\documentclass[10pt,twocolumn,superscriptaddress,showpacs,prl]{revtex4-1}  
\usepackage{amsmath}
\usepackage{verbatim}
\usepackage{graphicx}
\usepackage{subfigure}
\usepackage{bm}

\begin{document}

\title{\large{Intrinsic noise and discrete-time processes}}

\author{Joseph~D. Challenger}
\email{jchallenger@unifi.it}
\author{Duccio Fanelli}
\email{duccio.fanelli@unifi.it}
\affiliation{Dipartimento di Fisica e Astronomia, Universit\`{a} degli Studi di Firenze and INFN, via Sansone 1, IT 50019 Sesto Fiorentino, Italy}
\author{Alan~J. McKane}
\email{alan.mckane@manchester.ac.uk}
\affiliation{Theoretical Physics Division, School of Physics and Astronomy, 
The University of Manchester, Manchester M13 9PL, UK}

\begin{abstract}
A general formalism is developed to construct a Markov chain model that converges to a one-dimensional map in the infinite population limit. Stochastic fluctuations are therefore internal to the system and not externally specified. For finite populations an approximate Gaussian scheme is devised to describe the stochastic fluctuations in the non-chaotic regime. More generally, the stochastic dynamics can be captured using a stochastic difference equation, derived through an approximation to the Markov chain. The scheme is demonstrated using the logistic map as a case study.
\end{abstract}
\maketitle

The study of population dynamics has a wide range of applications. In cells, enzymes (proteins) catalyze reactions, interact with each other and with cell products, and regulate the synthesis of other proteins. On a larger scale, one faces the problem of understanding the behavior of bacterial communities under various environmental conditions and, at even larger scales, how ecological communities are assembled and epidemics controlled. Such populations of microscopic actors can be modeled as macroscopic abundances or concentrations. Population dynamics is often studied in this mean-field, deterministic, description. In contrast to this idealized picture, an individual-based model recognizes the inherent discreteness of the population~\cite{black_12}. Stochastic effects, stemming from the finite populations of interacting elements, can dramatically impact the behavior of dynamical systems. This is a widespread observation and has many applications, ranging from biology to physics.

For a continuous time process, an individual-based model can typically be described by a master equation~\cite{kampen_07}. In general, this equation cannot be solved exactly. However, approximate analytical methods exist to transform it into either a Fokker-Planck equation (FPE) or a stochastic differential equation~\cite{gardiner_04}. Under specific conditions, the stochasticity can be amplified, producing macroscopic order, such as sustained oscillations~\cite{mckane_05}. Finite population-size corrections also impact on systems for which the mean-field equations are chaotic. This important problem has so far only been addressed numerically, see e.g. Refs.~\cite{guemez_93,li_98}. 

Alternatively, discrete time maps can be used to investigate the onset of chaos. These simple deterministic systems have many applications e.g. the study of biological populations where successive generations do not overlap \cite{may_76}. From the 1980s onwards, the effects of noise on 1D maps has been extensively investigated~\cite{crutchfield_82,crutchfield_83,fraser_83,fox_90,gao_99,boccaletti_02,fogedby_05}. However, in all of these studies noise was simply added to the deterministic dynamics. This may be appropriate for the case of external noise, which is imposed on the system as a modeling choice. Systems such as the logistic map are however bounded on a finite interval: if the map is perturbed by e.g. an additive Gaussian noise, it is possible for a stochastic trajectory to escape this interval. One must then either impose an artificial constraint to prevent such an escape, or only explore the weak noise regime. By contrast, intrinsic noise has a structure appropriate to the system under consideration, which allows the strong noise regime to be explored.

In this Rapid Communication we develop a methodology for understanding the effects of intrinsic noise in 1D maps. We begin by showing how to construct a Markov chain, describing transitions between discrete states in discrete time intervals, which has a 1D map as its deterministic limit. %This is the first main result of this Letter. 
The Markov chain can be described mesoscopically by a generalization of the FPE or, equivalently, by a stochastic difference equation. A Gaussian closure leads to an approximate 2D map, characterizing the probability distribution. %These systematic theoretical characterizations of the intrinsic fluctuations constitute the second main result of this Letter. 
To develop the methodology, we take the logistic map as a case study. Where the map is non-chaotic, and for sufficiently large population sizes, fluctuations are described by the Gaussian approximation. In the strong noise regime, non-Gaussian effects are captured by the stochastic difference equation, as confirmed by direct comparison with the Markov chain simulations. The effect of the intrinsic noise is to destroy the periodicity of cyclic solutions and to anticipate the edge of chaos.
   
The starting point of our discussion is a 1D map which has the form $z_{t+1} = f (z_t)$. Here $z_t$ is a continuous variable, time $t$ changes in discrete steps and $f(\cdot)$ is a generic, nonlinear function. Let us begin by describing the general Markov chain model to be used in this work  and show how it is related to this deterministic map. We consider a population of $N$ individuals which at (discrete) time $t$ consists of $m$ of one type ($A$) and $(N-m)$ of another type ($B$). We then create a new population at (discrete) time $t+1$ by randomly sampling this original population. A matrix with entries $Q_{nm}$ specifies the sampling process, with $Q_{nm}$ giving the probability that there are $n$ $A$ individuals present at time $t+1$ given there were $m$ $A$ individuals present at time $t$. In other words, the $(N+1) \times (N+1)$ matrix $\bm{Q}$ specifies the probability of creating generation $(t+1)$ with a certain composition from generation $t$. To study the time evolution of the system, we specify its state at time $t$ by the vector $\bm{P}_{t}=(P_{0,t},P_{1,t},\hdots,P_{N,t})$, where $P_{n,t}$ is the probability for there to be $n$ individuals of type $A$ present at time $t$. The system then evolves in time according to 
\begin{equation}
\bm{P}_{t+1}=\bm{Q}\bm{P}_{t}.
\label{Markov_chain}
\end{equation}
The columns of $\bm{Q}$ each sum to unity and long-ranged transitions are allowed, a key requirement to obtain stochastic analogs of nonlinear deterministic maps. The particular form of $\bm{Q}$ we have chosen is inspired by the Wright-Fisher model of genetic drift~\cite{fisher_30,wright_31}, designed to describe a haploid population of size $N$, where the individuals have a gene which has two alleles ($A$ and $B$). We postulate the following general form for $\bm{Q}$,
\begin{equation}
\label{Qnm}
Q_{nm}= { N \choose n} \left[ f\left( \frac{m}{N}\right) \right]^{n}\,\left[ 1 - 
f\left( \frac{m}{N} \right) \right]^{N-n}.
\end{equation}
We require $f$ is such that $0 \leq f(m/N) \leq 1$ and ask that in the limit $N \to \infty$ it gives the nonlinear function of the deterministic map. We note here that the choice $f(m/N)=m/N$ recovers the original Wright-Fisher model. 

To make contact with this mean-field limit, we define the expectation value $\langle n \rangle = \sum_n n P_n$ and using Eq.~({\ref{Markov_chain}) find that
\begin{equation}
\langle n_{t+1} \rangle = \sum_n n P_{n,t+1} = \sum_n \sum_m n Q_{nm} P_{m,t}.
\end{equation}
By the definition of $Q_{nm}$ given in Eq.~(\ref{Qnm}), one readily finds $\sum_n n Q_{nm}= Nf(m/N)$, and therefore: 
\begin{equation}
\label{mf}
\langle n_{t+1} \rangle = \sum_m Nf\left(\frac{m}{N}\right) P_{m,t} = N\left\langle f\left(\frac{n_t}{N}\right) \right\rangle. 
\end{equation}
The deterministic limit is found by dividing both sides of Eq.~(\ref{mf}) by $N$ and taking the limit $N\to \infty$. In this limit $\langle f(n/N) \rangle = f(\langle n/N \rangle)$, so providing that the function $f$ is chosen appropriately, Eq.~(\ref{mf}) takes the form $z_{t+1}=f(z_t)$, where the variable $z_t=\lim_{N\to \infty} \langle n_t \rangle /N$ has been introduced. In summary, given a 1D map, we can define a matrix $\bm{Q}$ of the Wright-Fisher type according to Eq.~(\ref{Qnm}), using the function $f(\cdot)$ to provide a stochastic version of the discrete map. The stochastic model is guaranteed by construction to converge to the corresponding 1D map in the limit of infinite population size. We should stress that this is only one way of creating such a stochastic model, albeit a very natural one; there are an infinity of stochastic models corresponding to any one deterministic model. In a continuous time process, governed by a master equation, one characterizes the process by transition rates, detailing individual events such as births, deaths or predation. In the discrete time process studied here, the state of the system is updated only at fixed intervals. Therefore, one must instead detail the process by which one generation is produced by the previous one. In applications one would start from the Markov chain model, introducing any non-linear features such as dynamical feedbacks, mutual competition and saturation.

Starting from the above formulation and performing the Kramers-Moyal expansion, we obtain a generalization of the FPE. This equation provides a complete description of the process, which may also be represented by an equivalent stochastic difference equation, analogous to the Langevin equation for continuous time systems. As before, the deterministic map is recovered in the limit $N\to\infty$. In the following, we shall not present the mathematical aspects of the formalism, leaving them to a forthcoming publication~\cite{challenger_xx}, but demonstrate the predictive power of the theory by making comparisons with direct simulations of the Markov chain model.  

To investigate the role played by intrinsic fluctuations, we will follow instead a more intuitive, approximate strategy and estimate the second moment of the probability distribution, via a simple Gaussian closure. We begin with the equation for the second moment
\begin{equation}
\langle n^2_{t+1} \rangle=\sum_n n^2 P_{n,t+1}=\sum_m \left[ \sum_n n^2 Q_{nm} \right]P_{m,t}.
\end{equation}
We can now use the binomial structure of $\bm{Q}$ to write
\begin{align}
\label{mom2}
\langle n^2_{t+1} \rangle &=\sum_m\left[ (Nf)^2 +Nf(1-f) \right] P_m \nonumber\\
&=N^2\langle f^2 \rangle +N\langle f \rangle -N\langle f^2 \rangle,
\end{align}
writing $f(m/N)=f$ for brevity. As $f$ is non-linear, higher order moments will appear on the right-hand side of Eq.~(\ref{mom2}). Therefore, a closure is necessary to simultaneously solve Eqs.~(\ref{mf}) and (\ref{mom2}). A Gaussian closure replaces the $q^{\rm th}$ moment with $\langle n^q \rangle_G$, where $q\geq 3$, by \cite{lafuerza_10}
\begin{equation}
\langle n^q \rangle_G=\langle n \rangle^q +\sum_{k=1}^{\left[\frac{q}{2}\right]}{ q \choose 2k}(2k-1)!!\langle n \rangle^{q-2k}\left[\langle n^2 \rangle -\langle n \rangle^2 \right]^k,
\label{eq:close}
\end{equation}
where the upper limit on the sum is the integer part of $q/2$. We use Eq.~(\ref{eq:close}) to close the equations for the first two moments and so obtain an estimate for the mean and the variance of the probability distribution. 

The formalism developed so far can be applied to a general 1D map. To illustrate these ideas further, we will examine the celebrated logistic map $z_{t+1}=\lambda z_t(1-z_t)$, where $\lambda$ is a parameter. For $0\leq \lambda \leq 4$, $z_t$ remains in the unit interval for all time. A natural interpretation for $z_t$ would be the fraction of individuals in an ecological community at time $t$, with the individuals being labeled as type $A$ and the vacancies being labeled as type $B$. As $\lambda$ is varied, the map displays a variety of behaviors \cite{strogatz_94}. For $1\leq \lambda \leq 3$ the map has a stable non-zero fixed point, then a cascade of period doubling bifurcations take place, up to the onset of chaos at $\lambda=3.56995\hdots$. Due to its simple form, the logistic map has been widely employed as a canonical model of chaos. The behavior of the logistic map yields the famous bifurcation diagram, which has become an icon of non-linear dynamics, shown in the top panel of Figure~\ref{fig:bif_and_fluct}. 

Based on the above scheme, we will consider a stochastic version of the logistic map, a microscopic process of the Wright-Fisher type, involving $N$ individuals, with $f(m/N)=\lambda (m/N) (1-m/N)$ in Eq.~(\ref{Qnm}). The boundary conditions are natural to the process, and are given in Eq.~(\ref{Qnm}) for this choice of $f$. This Markov chain converges to the logistic map in the limit $N\to \infty$. At finite $N$, the evolution of the first moment $\langle n\rangle$ is governed by Eq.~(\ref{mf}). Since $f$ is quadratic, this latter equation involves the second moment $\langle n^2\rangle$. To proceed, we insert the explicit form for $f$ into Eq.~(\ref{mom2}), the equation for the second moment. Using the Gaussian closure in (\ref{eq:close}) we find
\begin{eqnarray}
\label{eq:gauss}
x_{t+1}&=&\lambda(x_{t}-y_{t}),\nonumber \\
y_{t+1}&=&\frac{\lambda}{N}\left[ x_t-y_t \right] \\ &+&(1-\frac{1}{N})\lambda^2\left[ y_t -2(3y_t x_t-2x_t^3)+3y_t^2-2x_t^4\right],\nonumber \\ 
\nonumber
\end{eqnarray}
where we write $x_t=\langle n_{t}\rangle/N$ and $y_t=\langle n^2_{t}\rangle/N^2$, and where $N$ is assumed to be large. In the case where $N$ is finite, we use $x_t$ to denote the (scaled) first moment, to distinguish from $z_t$, used in the mean-field limit. These coupled equations approximate the stochastic dynamics of the Markov chain. We will start by showing results obtained with this approximation and then compare them with simulations. 

The stochastic difference equation, obtained from the generalization of the FPE~\cite{challenger_xx}, can also be utilized to characterize the stochastic dynamics. For the case of the logistic map, we find that the variable $z$, now a random variable, obeys the equation $z_{t+1}=\lambda z_t(1-z_t)+\eta_t$, where $\eta_t$ is a Gaussian variable with zero mean and correlator $\langle \eta_t\eta_{t'}\rangle=[\lambda z_{t}(1-z_{t})(1-\lambda z_{t}(1-z_{t}))/N]\delta_{t\,t'}$. The noise is multiplicative, in contrast to the additive Gaussian noise proposed as an external perturbation in previous studies.

\begin{figure}
\subfigure{\label{fig:label1}\includegraphics[scale=0.65]{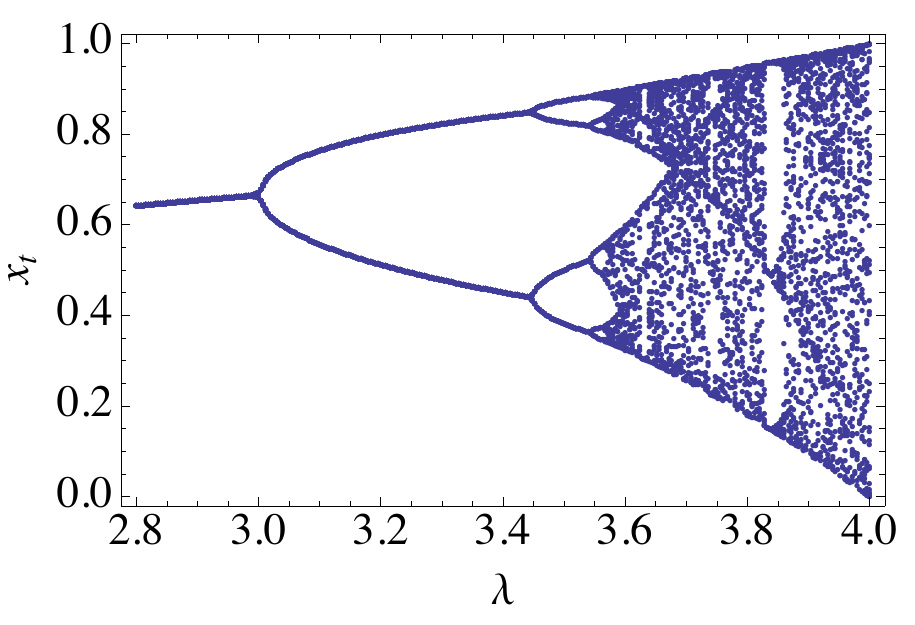}}
\subfigure{\label{fig:label3}\includegraphics[scale=0.65]{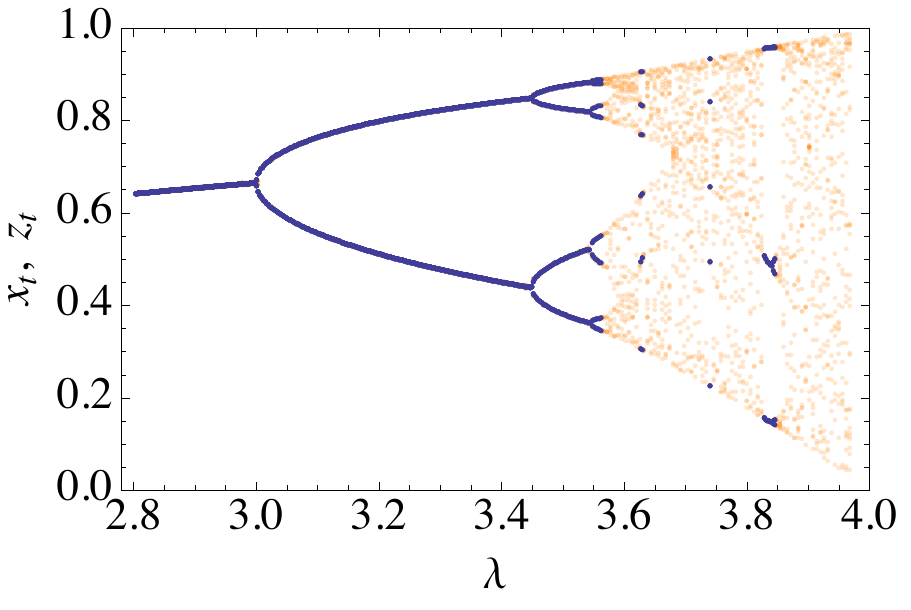}}
\subfigure{\label{fig:label5}\includegraphics[scale=0.65]{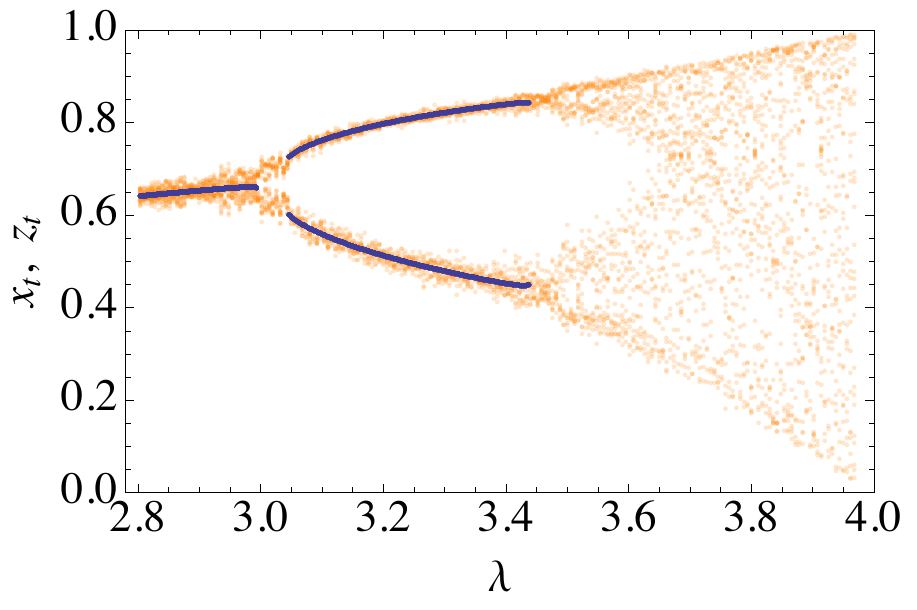}}
\caption{
Bifurcation diagrams for different values of $N$. The top panel shows the deterministic logistic map. The middle and lower panels show the Gaussian approximation for the first moment, $x_t$, for $N=5\times 10^6$ and $N=5000$ respectively (blue lines). These diagrams show values of $\lambda$ where a solution of the Gaussian approximation was found. Simulation results from the stochastic difference equation (orange dots) are also shown.
}
\label{fig:bif_and_fluct}
\end{figure}

We begin by depicting the bifurcation diagram for $x_t$ for different $N$, generated by iterating Eq.~(\ref{eq:gauss}) from an initial condition, for a given value of $\lambda$. After the transient has died out, the values of $x_t$ are plotted against the selected $\lambda$. A fixed point appears in the diagram as a single spot, whilst an $s$-cycle appears as $s$ distinct points. The results are shown in Figure~\ref{fig:bif_and_fluct} for different $N$. When $N$ is reduced, portions of the diagrams fade away, due to the loss of stability of the 2D map. It is found that stability is lost when  the fluctuations are large enough that the basins of attraction associated with each branch of the asymptotic solution are no longer isolated. In other words, if $\lambda$ has a value for which the deterministic system has an $s$-cycle, the stationary probability distribution may be approximated by a superposition of $s$ Gaussians. For $N$ sufficiently large, the peaks of the Gaussians are well separated, and correspond to the $s$-cycles of the deterministic map. By reducing $N$, the widths of the Gaussians grow, and can overlap, and the system loses the periodicity associated with the deterministic $s$-cycle. The complete, finite $N$, bifurcation diagram can be reconstructed from the stochastic difference equation for the random variable $z$, as displayed in the lower panels of Figure~\ref{fig:bif_and_fluct} (orange dots). This enables us to see how the system behaves when the Gaussian approximation fails, and also appreciate the strength of the fluctuations around the mean, $x_t$. We simulate the stochastic difference equation here, rather than the original Markov chain, as for the former the computation time is independent of $N$. 

\begin{figure}
\subfigure{
 \includegraphics[scale=0.91]{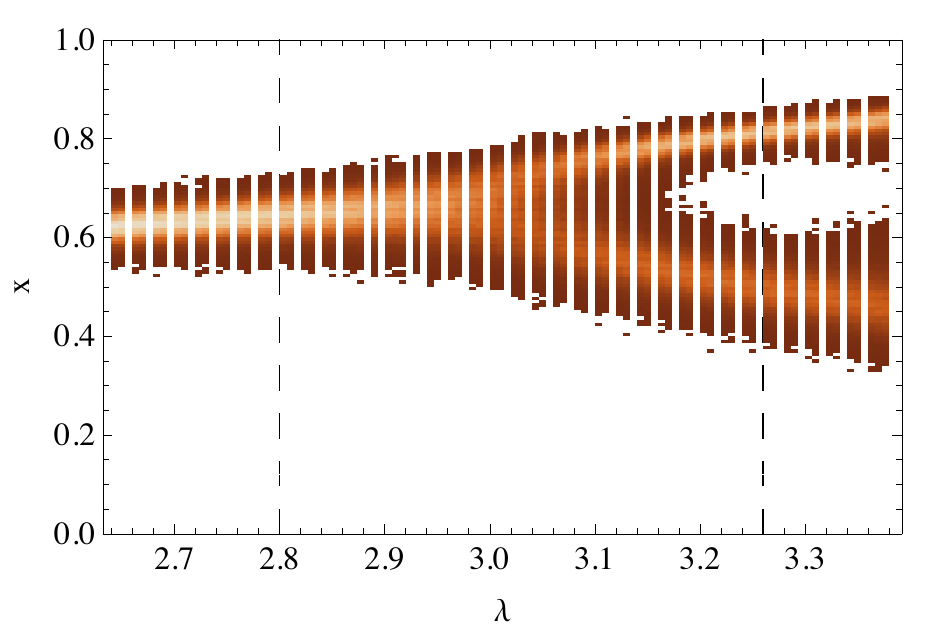}
}
\\
 \begin{center}
  \subfigure{\label{fig:quelquechose1}\includegraphics[scale=0.46]{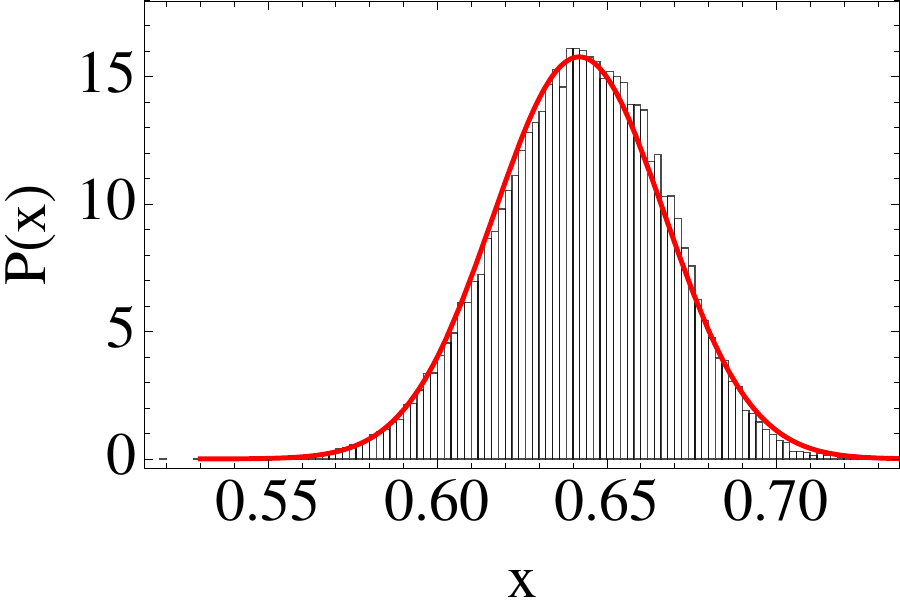}}
\subfigure{\label{fig:quelquechose2}\includegraphics[scale=0.46]{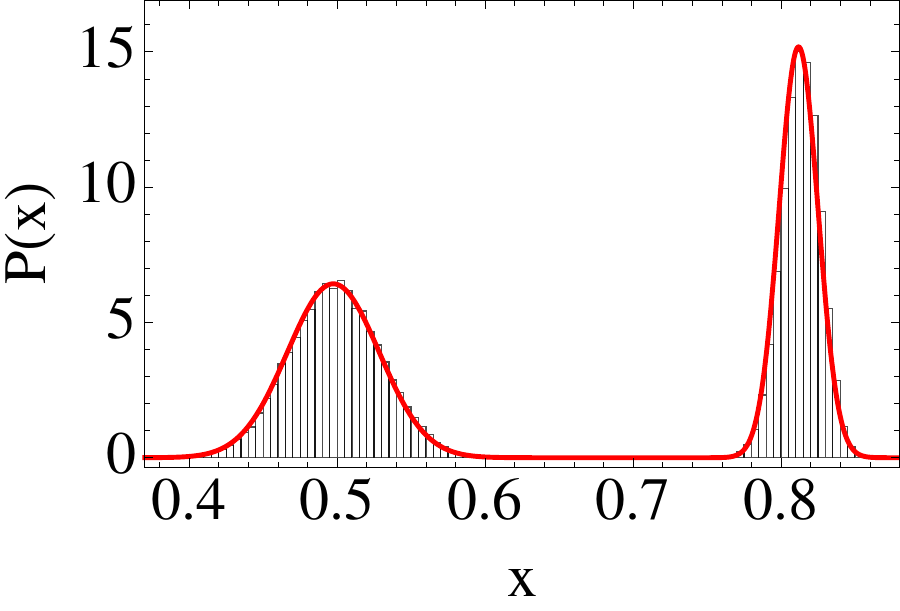}}
 \end{center}
\caption{Above: bifurcation diagram generated by simulating the Markov chain for $N=1000$. For each value of $\lambda$ the chain was iterated 20000 times. The color denotes the relative frequency with which the states were visited. Below: distributions for this Markov chain. The values of $\lambda$ chosen are $\lambda=2.8$ (left) and $\lambda=3.26$ (right), denoted by the dashed lines in the top panel. The bars refer to stationary distributions obtained by simulating the chain. The red lines show the Gaussian approximation.}
\label{fig:pair1}
\end{figure}

\begin{figure}[h]
\begin{center}
\includegraphics[scale=0.86]{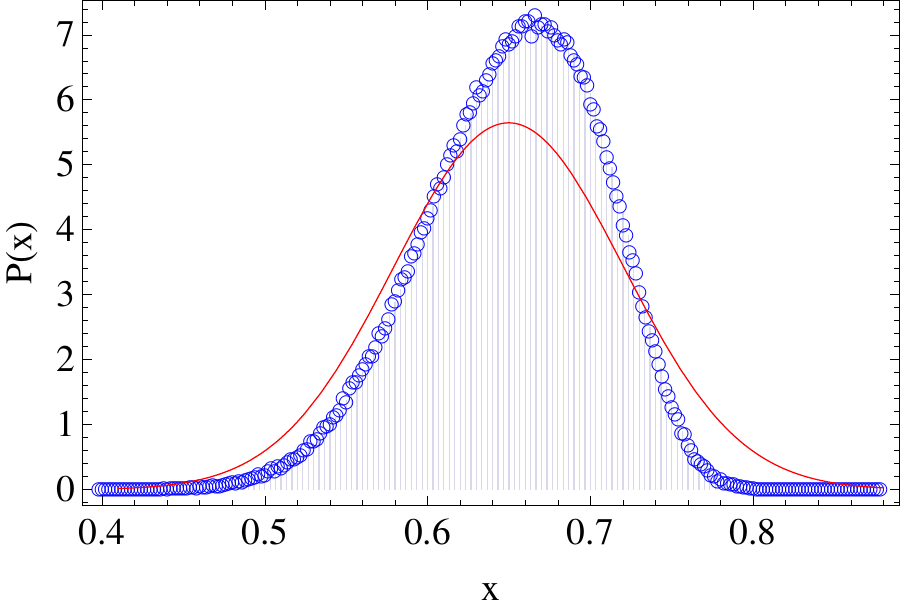}
\end{center}
\caption{The probability distribution function for a case where the Gaussian approximation loses accuracy. The stationary distribution found from the Markov chain (bars) is well captured by simulations of the corresponding stochastic difference equation (blue circles), but not by the approximate 2D map (red line). Parameter values are $N=300$ and $\lambda=2.92$.}
\label{fig:3curves}
\end{figure}

We now turn to simulate the Markov chain to test the accuracy of the theoretical schemes proposed. The upper part of Figure~\ref{fig:pair1} shows a section of a bifurcation diagram, generated by simulating the Markov chain with $N=1000$, over a range of $\lambda$. Lighter patches identify more frequently visited states. In the lower panels of Figure~\ref{fig:pair1}, the probability distribution generated from the simulations is compared to the predictions from the Gaussian approximation for two choices of $\lambda$. In both cases, the results show excellent agreement with the simulations. The Gaussian prediction starts to lose accuracy when $N$ is small, or when $\lambda$ approaches a bifurcation point. Non-Gaussian corrections can  be calculated from the previously mentioned stochastic difference equation~\cite{challenger_xx}, as we show in Figure~\ref{fig:3curves}. For the parameters chosen here, the stationary distribution of the Markov chain is noticeably skewed. The distribution obtained from the stochastic difference equation is in excellent agreement with that found from the Markov chain.

Previous numerical studies have investigated the interplay between additive noise and chaos \cite{crutchfield_82}. We examine this phenomenon for our intrinsic noise process, focusing on how the noise affects the onset of chaos. We consider a 3-cycle, which appears as a regular window beyond the onset of chaos, for values of $\lambda$ around 3.83. Crutchfield et al.~\cite{crutchfield_82} showed that sufficiently strong additive noise can destroy the 3-cycle and, instead of three narrow peaks, one sees a broad probability distribution, akin to the type of distribution found for a choice of $\lambda$ for which the dynamics are chaotic. This is consistent with Figure~\ref{fig:bif_and_fluct}, where the Gaussian scheme can calculate fluctuations around the 3-cycle in the middle panel (for $N=5\times 10^6$) but not in the lower panel (where $N=5000$). The results for our intrinsic noise system are qualitatively similar to those found in \cite{crutchfield_82}: the Lyapunov exponent can be measured as a function of $N$ (data not shown) and a noise-induced transition to chaos is observed. One advantage of using an intrinsic noise process to investigate this question is that one has access to the strong noise limit, which can be difficult to explore using an additive noise process which is not characteristic of the system. 

To conlcude, we have developed a general strategy to construct a Markov chain which converges to a map in the infinite $N$ limit. Intrinsic fluctuations are approximated by a 2D map for the first two moments of the probability density distribution. More generally, a stochastic difference equation, reminiscent of the Langevin equation for continuous time processes, can be obtained and provides an excellent effective description of the Markov chain dynamics, even in the strong noise regime, where the Gaussian approximation breaks down. Results are shown here for the logistic map. However, the formalism is general and applies in all contexts, from physics to biology, where maps are employed as model systems.

JDC and DF acknowledge support from Program Prin2009 financed by the Italian MIUR.

%

%\bibliography{literature}

\begin{thebibliography}{22}%
\makeatletter
\providecommand \@ifxundefined [1]{%
 \@ifx{#1\undefined}
}%
\providecommand \@ifnum [1]{%
 \ifnum #1\expandafter \@firstoftwo
 \else \expandafter \@secondoftwo
 \fi
}%
\providecommand \@ifx [1]{%
 \ifx #1\expandafter \@firstoftwo
 \else \expandafter \@secondoftwo
 \fi
}%
\providecommand \natexlab [1]{#1}%
\providecommand \enquote  [1]{``#1''}%
\providecommand \bibnamefont  [1]{#1}%
\providecommand \bibfnamefont [1]{#1}%
\providecommand \citenamefont [1]{#1}%
\providecommand \href@noop [0]{\@secondoftwo}%
\providecommand \href [0]{\begingroup \@sanitize@url \@href}%
\providecommand \@href[1]{\@@startlink{#1}\@@href}%
\providecommand \@@href[1]{\endgroup#1\@@endlink}%
\providecommand \@sanitize@url [0]{\catcode `\\12\catcode `\$12\catcode
  `\&12\catcode `\#12\catcode `\^12\catcode `\_12\catcode `\%12\relax}%
\providecommand \@@startlink[1]{}%
\providecommand \@@endlink[0]{}%
\providecommand \url  [0]{\begingroup\@sanitize@url \@url }%
\providecommand \@url [1]{\endgroup\@href {#1}{\urlprefix }}%
\providecommand \urlprefix  [0]{URL }%
\providecommand \Eprint [0]{\href }%
\providecommand \doibase [0]{http://dx.doi.org/}%
\providecommand \selectlanguage [0]{\@gobble}%
\providecommand \bibinfo  [0]{\@secondoftwo}%
\providecommand \bibfield  [0]{\@secondoftwo}%
\providecommand \translation [1]{[#1]}%
\providecommand \BibitemOpen [0]{}%
\providecommand \bibitemStop [0]{}%
\providecommand \bibitemNoStop [0]{.\EOS\space}%
\providecommand \EOS [0]{\spacefactor3000\relax}%
\providecommand \BibitemShut  [1]{\csname bibitem#1\endcsname}%
\let\auto@bib@innerbib\@empty
%</preamble>
\bibitem [{\citenamefont {Black}\ and\ \citenamefont
  {McKane}(2012)}]{black_12}%
  \BibitemOpen
  \bibfield  {author} {\bibinfo {author} {\bibfnamefont {A.~J.}\ \bibnamefont
  {Black}}\ and\ \bibinfo {author} {\bibfnamefont {A.~J.}\ \bibnamefont
  {McKane}},\ }\href@noop {} {\bibfield  {journal} {\bibinfo  {journal} {Trends
  in Ecology \& Evolution}\ }\textbf {\bibinfo {volume} {27}},\ \bibinfo
  {pages} {337} (\bibinfo {year} {2012})}\BibitemShut {NoStop}%
\bibitem [{\citenamefont {{van Kampen}}(2007)}]{kampen_07}%
  \BibitemOpen
  \bibfield  {author} {\bibinfo {author} {\bibfnamefont {N.~G.}\ \bibnamefont
  {{van Kampen}}},\ }\href@noop {} {\emph {\bibinfo {title} {Stochastic
  Processes in Physics and Chemistry}}},\ \bibinfo {edition} {3rd}\ ed.\
  (\bibinfo  {publisher} {Elsevier},\ \bibinfo {address} {Amsterdam},\ \bibinfo
  {year} {2007})\BibitemShut {NoStop}%
\bibitem [{\citenamefont {Gardiner}(2009)}]{gardiner_04}%
  \BibitemOpen
  \bibfield  {author} {\bibinfo {author} {\bibfnamefont {C.~W.}\ \bibnamefont
  {Gardiner}},\ }\href@noop {} {\emph {\bibinfo {title} {Handbook of Stochastic
  Methods}}},\ \bibinfo {edition} {4th}\ ed.\ (\bibinfo  {publisher}
  {Springer-Verlag},\ \bibinfo {address} {Berlin},\ \bibinfo {year}
  {2009})\BibitemShut {NoStop}%
\bibitem [{\citenamefont {McKane}\ and\ \citenamefont
  {Newman}(2005)}]{mckane_05}%
  \BibitemOpen
  \bibfield  {author} {\bibinfo {author} {\bibfnamefont {A.~J.}\ \bibnamefont
  {McKane}}\ and\ \bibinfo {author} {\bibfnamefont {T.~J.}\ \bibnamefont
  {Newman}},\ }\href@noop {} {\bibfield  {journal} {\bibinfo  {journal} {Phys.
  Rev. Lett.}\ }\textbf {\bibinfo {volume} {94}},\ \bibinfo {pages} {218102}
  (\bibinfo {year} {2005})}\BibitemShut {NoStop}%
\bibitem [{\citenamefont {{G\"{u}\'{e}mez}}\ and\ \citenamefont
  {{Mat\'{i}as}}(1993)}]{guemez_93}%
  \BibitemOpen
  \bibfield  {author} {\bibinfo {author} {\bibfnamefont {J.}~\bibnamefont
  {{G\"{u}\'{e}mez}}}\ and\ \bibinfo {author} {\bibfnamefont {M.~A.}\
  \bibnamefont {{Mat\'{i}as}}},\ }\href@noop {} {\bibfield  {journal} {\bibinfo
   {journal} {Phys. Rev. E}\ }\textbf {\bibinfo {volume} {48}},\ \bibinfo
  {pages} {R2351} (\bibinfo {year} {1993})}\BibitemShut {NoStop}%
\bibitem [{\citenamefont {Li}\ and\ \citenamefont {Wang}(1998)}]{li_98}%
  \BibitemOpen
  \bibfield  {author} {\bibinfo {author} {\bibfnamefont {Q.}~\bibnamefont
  {Li}}\ and\ \bibinfo {author} {\bibfnamefont {H.}~\bibnamefont {Wang}},\
  }\href@noop {} {\bibfield  {journal} {\bibinfo  {journal} {Phys. Rev. E}\
  }\textbf {\bibinfo {volume} {58}},\ \bibinfo {pages} {R1191} (\bibinfo {year}
  {1998})}\BibitemShut {NoStop}%
\bibitem [{\citenamefont {May}(1976)}]{may_76}%
  \BibitemOpen
  \bibfield  {author} {\bibinfo {author} {\bibfnamefont {R.~M.}\ \bibnamefont
  {May}},\ }\href@noop {} {\bibfield  {journal} {\bibinfo  {journal} {Nature}\
  }\textbf {\bibinfo {volume} {261}},\ \bibinfo {pages} {459} (\bibinfo {year}
  {1976})}\BibitemShut {NoStop}%
\bibitem [{\citenamefont {Crutchfield}\ \emph {et~al.}(1982)\citenamefont
  {Crutchfield}, \citenamefont {Farmer},\ and\ \citenamefont
  {Huberman}}]{crutchfield_82}%
  \BibitemOpen
  \bibfield  {author} {\bibinfo {author} {\bibfnamefont {J.~P.}\ \bibnamefont
  {Crutchfield}}, \bibinfo {author} {\bibfnamefont {J.~D.}\ \bibnamefont
  {Farmer}}, \ and\ \bibinfo {author} {\bibfnamefont {B.~A.}\ \bibnamefont
  {Huberman}},\ }\href@noop {} {\bibfield  {journal} {\bibinfo  {journal}
  {Physics Reports}\ }\textbf {\bibinfo {volume} {92}},\ \bibinfo {pages} {45}
  (\bibinfo {year} {1982})}\BibitemShut {NoStop}%
\bibitem [{\citenamefont {Crutchfield}\ and\ \citenamefont
  {Packard}(1983)}]{crutchfield_83}%
  \BibitemOpen
  \bibfield  {author} {\bibinfo {author} {\bibfnamefont {J.~P.}\ \bibnamefont
  {Crutchfield}}\ and\ \bibinfo {author} {\bibfnamefont {N.~H.}\ \bibnamefont
  {Packard}},\ }\href@noop {} {\bibfield  {journal} {\bibinfo  {journal}
  {Physica D}\ }\textbf {\bibinfo {volume} {7}},\ \bibinfo {pages} {201}
  (\bibinfo {year} {1983})}\BibitemShut {NoStop}%
\bibitem [{\citenamefont {Fraser}\ \emph {et~al.}(1983)\citenamefont {Fraser},
  \citenamefont {Celarier},\ and\ \citenamefont {Kapral}}]{fraser_83}%
  \BibitemOpen
  \bibfield  {author} {\bibinfo {author} {\bibfnamefont {S.}~\bibnamefont
  {Fraser}}, \bibinfo {author} {\bibfnamefont {E.}~\bibnamefont {Celarier}}, \
  and\ \bibinfo {author} {\bibfnamefont {R.}~\bibnamefont {Kapral}},\
  }\href@noop {} {\bibfield  {journal} {\bibinfo  {journal} {J. Stat. Phys}\
  }\textbf {\bibinfo {volume} {33}},\ \bibinfo {pages} {341} (\bibinfo {year}
  {1983})}\BibitemShut {NoStop}%
\bibitem [{\citenamefont {Fox}(1990)}]{fox_90}%
  \BibitemOpen
  \bibfield  {author} {\bibinfo {author} {\bibfnamefont {R.~F.}\ \bibnamefont
  {Fox}},\ }\href@noop {} {\bibfield  {journal} {\bibinfo  {journal} {Phys.
  Rev. A}\ }\textbf {\bibinfo {volume} {42}},\ \bibinfo {pages} {1946}
  (\bibinfo {year} {1990})}\BibitemShut {NoStop}%
\bibitem [{\citenamefont {Gao}\ \emph {et~al.}(1999)\citenamefont {Gao},
  \citenamefont {Hwang},\ and\ \citenamefont {Liu}}]{gao_99}%
  \BibitemOpen
  \bibfield  {author} {\bibinfo {author} {\bibfnamefont {J.~B.}\ \bibnamefont
  {Gao}}, \bibinfo {author} {\bibfnamefont {S.~K.}\ \bibnamefont {Hwang}}, \
  and\ \bibinfo {author} {\bibfnamefont {J.~M.}\ \bibnamefont {Liu}},\
  }\href@noop {} {\bibfield  {journal} {\bibinfo  {journal} {Phys. Rev. Lett.}\
  }\textbf {\bibinfo {volume} {82}},\ \bibinfo {pages} {1132} (\bibinfo {year}
  {1999})}\BibitemShut {NoStop}%
\bibitem [{\citenamefont {Boccaletti}\ \emph {et~al.}(2002)\citenamefont
  {Boccaletti}, \citenamefont {Kurths}, \citenamefont {Osipov}, \citenamefont
  {Valladares},\ and\ \citenamefont {Zhou}}]{boccaletti_02}%
  \BibitemOpen
  \bibfield  {author} {\bibinfo {author} {\bibfnamefont {S.}~\bibnamefont
  {Boccaletti}}, \bibinfo {author} {\bibfnamefont {J.}~\bibnamefont {Kurths}},
  \bibinfo {author} {\bibfnamefont {G.}~\bibnamefont {Osipov}}, \bibinfo
  {author} {\bibfnamefont {D.~L.}\ \bibnamefont {Valladares}}, \ and\ \bibinfo
  {author} {\bibfnamefont {C.~S.}\ \bibnamefont {Zhou}},\ }\href@noop {}
  {\bibfield  {journal} {\bibinfo  {journal} {Physics Reports}\ }\textbf
  {\bibinfo {volume} {366}},\ \bibinfo {pages} {1} (\bibinfo {year}
  {2002})}\BibitemShut {NoStop}%
\bibitem [{\citenamefont {Fogedby}\ and\ \citenamefont
  {Jensen}(2005)}]{fogedby_05}%
  \BibitemOpen
  \bibfield  {author} {\bibinfo {author} {\bibfnamefont {H.~C.}\ \bibnamefont
  {Fogedby}}\ and\ \bibinfo {author} {\bibfnamefont {M.~H.}\ \bibnamefont
  {Jensen}},\ }\href@noop {} {\bibfield  {journal} {\bibinfo  {journal} {J.
  Stat. Phys.}\ }\textbf {\bibinfo {volume} {121}},\ \bibinfo {pages} {759}
  (\bibinfo {year} {2005})}\BibitemShut {NoStop}%
\bibitem [{\citenamefont {Fisher}(1930)}]{fisher_30}%
  \BibitemOpen
  \bibfield  {author} {\bibinfo {author} {\bibfnamefont {R.~A.}\ \bibnamefont
  {Fisher}},\ }\href@noop {} {\emph {\bibinfo {title} {The Genetical Theory of
  Natural Selection}}}\ (\bibinfo  {publisher} {Clarendon},\ \bibinfo {address}
  {Oxford},\ \bibinfo {year} {1930})\BibitemShut {NoStop}%
\bibitem [{\citenamefont {Wright}(1931)}]{wright_31}%
  \BibitemOpen
  \bibfield  {author} {\bibinfo {author} {\bibfnamefont {S.}~\bibnamefont
  {Wright}},\ }\href@noop {} {\bibfield  {journal} {\bibinfo  {journal}
  {Genetics}\ }\textbf {\bibinfo {volume} {16}},\ \bibinfo {pages} {97}
  (\bibinfo {year} {1931})}\BibitemShut {NoStop}%
\bibitem [{\citenamefont {Challenger}\ \emph {et~al.}()\citenamefont
  {Challenger}, \citenamefont {Fanelli},\ and\ \citenamefont
  {McKane}}]{challenger_xx}%
  \BibitemOpen
  \bibfield  {author} {\bibinfo {author} {\bibfnamefont {J.~D.}\ \bibnamefont
  {Challenger}}, \bibinfo {author} {\bibfnamefont {D.}~\bibnamefont {Fanelli}},
  \ and\ \bibinfo {author} {\bibfnamefont {A.~J.}\ \bibnamefont {McKane}},\
  }\href@noop {} {}\bibinfo {note} {To be published}\BibitemShut {NoStop}%
\bibitem [{\citenamefont {Lafuerza}\ and\ \citenamefont
  {Toral}(2010)}]{lafuerza_10}%
  \BibitemOpen
  \bibfield  {author} {\bibinfo {author} {\bibfnamefont {L.~F.}\ \bibnamefont
  {Lafuerza}}\ and\ \bibinfo {author} {\bibfnamefont {R.}~\bibnamefont
  {Toral}},\ }\href@noop {} {\bibfield  {journal} {\bibinfo  {journal} {J.
  Stat. Phys.}\ }\textbf {\bibinfo {volume} {140}},\ \bibinfo {pages} {917}
  (\bibinfo {year} {2010})}\BibitemShut {NoStop}%
\bibitem [{\citenamefont {Strogatz}(1994)}]{strogatz_94}%
  \BibitemOpen
  \bibfield  {author} {\bibinfo {author} {\bibfnamefont {S.~H.}\ \bibnamefont
  {Strogatz}},\ }\href@noop {} {\emph {\bibinfo {title} {Nonlinear Dynamics and
  Chaos}}}\ (\bibinfo  {publisher} {Perseus Books},\ \bibinfo {address}
  {Cambridge, MA},\ \bibinfo {year} {1994})\BibitemShut {NoStop}%
\end{thebibliography}
\end{document}